\documentclass{article}

 \UseRawInputEncoding 

\newtheorem{theorem}{Theorem}[section]
\newtheorem{lemma}[theorem]{Lemma}
\newtheorem{corollary}[theorem]{Corollary}
\newtheorem{definition}[theorem]{Definition}
\newtheorem{problem}[theorem]{Problem}

\newcommand\qed{\begin{flushright} {\bf q.e.d.} \end{flushright} }
\newcommand\prf{\noindent {\bf Proof :}}  
\newcommand\aprf{\noindent {\bf First proof :}}
\newcommand\bprf{\noindent {\bf Second proof :}}

\newcommand\bits{\{0,1\}}
\newcommand\uu{{\bits^*}}
\newcommand\nn{{\bits^n}}
\newcommand\mm{{\bits^m}}

\newcommand\tru{{\mbox{{\bf tt}}}}

\begin{document}

\title{Information in propositional proofs\\
and algorithmic proof search}  

\author{Jan Kraj\'{\i}\v{c}ek}

\date{Faculty of Mathematics and Physics\\
Charles University\thanks{
Sokolovsk\' a 83, Prague, 186 75,
The Czech Republic,
{\tt krajicek@karlin.mff.cuni.cz}}}

\maketitle

\begin{abstract}
We study from the proof complexity perspective
the (informal) proof search problem (cf. \cite[Secs.1.5 and 21.5]{prf}):

\begin{itemize}

\item {\em Is there an optimal way to search for propositional proofs?
}
\end{itemize}
\noindent
We note that, as a consequence of Levin's universal search,
for any fixed proof system there exists a time-optimal proof search algorithm. 
Using classical proof complexity results about reflection principles we prove 
that a time-optimal proof search algorithm exists without
restricting proof systems iff a p-optimal proof system exists.

To characterize precisely the time proof search algorithms need for individual formulas
we introduce a new proof complexity
measure based on algorithmic information concepts. In particular, to a proof system $P$ we attach
{\bf information-efficiency function} $i_P(\tau)$ assigning to a tautology a natural number, and we show
that:
\begin{itemize}

\item $i_P(\tau)$ characterizes time any $P$-proof search algorithm has to use on $\tau$,

\item for a fixed $P$ there is such an information-optimal algorithm (informally: it finds proofs 
of minimal information content),

\item a proof system is information-efficiency optimal (its information-efficiency function
is minimal up to a multiplicative constant)
iff it is p-optimal,

\item for non-automatizable systems $P$ there are formulas $\tau$ with short proofs but having large information measure 
$i_P(\tau)$.

\end{itemize}
We isolate and motivate the problem to establish {\em unconditional} super-logarithmic lower bounds for $i_P(\tau)$
where no super-polynomial size lower bounds are known.
We also point out connections of the new measure with some topics in proof complexity other than proof search. 
\end{abstract}

\section{Introduction}
The central notion of proof complexity is that of
a {\bf propositional proof system} as defined by Cook and Reckhow \cite{CooRec}: a p-time function
$$
P\ :\ \uu \rightarrow \uu
$$
whose range is exactly the set of propositional tautologies TAUT; for the definiteness we take 
all tautologies in the DeMorgan language. Any $w \in \uu$ such that $P(w) = \tau$ is called
a {\bf $P$-proof of $\tau$}. 

The primary concern is the size of proofs (i.e. their bit-length)
and, in particular, the existence of short proofs. The efficiency of a proof system $P$
is measured by the growth rate of {\bf the length-of-proof function}:
$$
s_P(\tau) := \min\{|w|\ |\ \mbox{ $w$ is a $P$-proof of $\tau$ }\}\ .
$$
(We are interested in values of these functions on TAUT only and may, for the definiteness, define
it to be equal to $\infty$ outside TAUT.)
 
{\bf The fundamental problem} of proof complexity theory asks if this function is, for some $P$, 
bounded by $|\tau|^{O(1)}$, for all $\tau \in \mbox{TAUT}$. 
This is equivalent to the question whether the computational complexity
class NP is closed under complementation: {\bf $\mbox{NP} =_? \mbox{coNP}$}, cf. Cook and Reckhow \cite{CooRec}.

The second principal open problem of proof complexity is {\bf the optimality problem}: Is there a 
proof system $P$ such that $s_P$ has at most polynomial slow-down over any $s_Q$? If we define
a quasi-ordering $P \geq Q$ on the set of all proof systems by
$$
s_P(\tau) \le s_Q(\tau)^{O(1)}
$$
then the problem asks if there is a $\geq$-maximal proof system.
Such a maximal proof system $P$ would be lengths-of-proofs optimal.
The quasi-ordering $\geq$ has a finer version $\geq_p$: $P \geq_p Q$ iff there is
a p-time function $f$ (called {\bf p-simulation}) such that for all $w$:
$$
P(f(w)) = Q(w)\ .
$$
In words: $f$ translates $Q$-proofs into $P$-proofs of the same formulas. 
The reader can find this basic background in \cite[Chpt.1]{prf}. 

\bigskip

While the existence of short proofs of tautologies is the primary concern of proof complexity,
the theory also relates quite closely to the complexity of proof search and SAT algorithms.
For proof search algorithms this is obvious: the time complexity of an algorithm searching for $P$-proofs 
is lower-bounded by function $s_P$. 

For SAT algorithms (i.e. algorithms finding a satisfying assignment for a propositional formula, if it exists)
the relation is indirect. 
In particular, we can interpret the run of a SAT algorithm $S$ that fails to find a satisfying assignment
for $\neg \tau$ as a {\em proof} that $\tau \in \mbox{TAUT}$. Hence $S$ can be studied also as a proof system $P_S$:
$$
P_S(w) = \tau\ \mbox{ iff }\ 
\mbox{({\em  $w$ is the failing computation of $S$ on $\neg \tau$}) }
$$
and the time complexity of $S$ on unsatisfiable formulas is essentially the same as the length-of-proof
function for proof system\footnote{Note that algorithm $S$ is, in particular, also an
algorithm searching for $P_S$-proofs.} $P_S$. Hence lower bounds to the latter function imply lower bounds for the time complexity of $S$.
In fact, proof complexity lower bounds apply more generally:
a lower bound for $s_Q$ implies time lower bounds for all SAT algorithms whose soundness is
efficiently provable\footnote{There are p-size $Q$-proofs of propositional translations of the first-order
statement formalizing the soundness.} in $Q$, cf. \cite{Kra-sat}.

While $P_S$ is defined more abstractly than usual logical calculi, the proof system is actually often equal (or close) 
- in the sense of p-simulation - to some
standard logical calculi as is, for example, {\bf resolution R}. This then allows to interpret
various technical proof complexity results as results about the original
algorithm $S$. In this sense proof complexity contributes to the analysis of some classes of SAT algorithms. This facet of 
proof complexity is surveyed by Buss and Nordstr\"{o}m \cite{BusNor}. 

\bigskip

In this paper we are interested in efficiency of proof search algorithms. 
However, rather than analyzing particular algorithms 
we consider an outstanding informal problem\footnote{Which we included as a third basic problem
of proof complexity under the name {\bf the proof search problem} in \cite[Secs.1.5 and 21.5]{prf}}:

\begin{itemize}

\item {\em Is there an optimal way to search for propositional proofs?
}
\end{itemize}
\noindent
Surely this problem must have occurred to everybody interested in proof search, and
there are other natural informal questions one can ask (cf. \cite{Kra-mfo} for examples).

In this paper we investigate what can proof complexity say about the problem in precise mathematical terms. 
We start with a definition 
of a proof search algorithm that seems natural (cf. \cite[Sec.21.5]{prf}).

\begin{definition} \label{27.2.21a}
A {\bf proof search algorithm} is a pair $(A,P)$, where $P$ is a proof system and $A$ is a deterministic
algorithm that stops on every input\footnote{See the second paragraph of Sec.\ref{1} for this condition.} 
and such that $A(\tau)$ is a $P$-proof of $\tau$, for all tautologies 
$\tau \in \mbox{TAUT}$. 
\end{definition}

A key to a formalization of the proof search problem is to
define a suitable quasi-ordering on the class of all proof search algorithms.
In Section \ref{1} we consider a quasi-ordering by the time complexity
and in Section \ref{2} we resort to algorithmic information theory 
and replace time with information, introducing a new  notion of
information-efficiency of proof systems. This notion offers a precise characterization of the time any
algorithm searching for a proof of a particular formula must use.

In both quasi-orderings (by time or information efficiency) 
there are optimal proof search algorithms when the proof system is fixed, and these two algorithms are essentially 
the same.
Hence the question whether there is an overall optimal proof search algorithm $(A,P)$ (a maximal element 
in the respective quasi-ordering) depends only on proof systems $P$ and not on algorithms $A$.
We show that in both quasi-orderings the existence of such an optimal system is equivalent to the existence
of a p-optimal proof system, and thus the proof search problem reduces to the optimality problem.

Time a proof search algorithm needs to use is traditionally lower bounded by the minimum size of
any proof of the formula in question. 
In Section \ref{3} we compare size (measure) with information
(measure) and we note that for non-automatizable proof systems 
the information measure is more precise for proving time lower bounds than proof size is: 
there are formulas having short proofs but having large information measure
(i.e. while the proofs are short to find them requires long time). Note that it is known that essentially all
complete proof systems are non-automatizable under various plausible computational complexity hypotheses.
In Section \ref{5.4.21a} we motivate and isolate the problem to establish {\em unconditional} lower bounds for
$i_P(\tau)$ where no lower bounds are known for $s_P(\tau)$.
We conclude the paper with remarks on several connections of the information measure to
proof complexity in Section \ref{4} and with some further comments in Section \ref{9.4.21c}.

\bigskip

The reader can find basic proof complexity background in \cite[Chpt.1]{prf}.
We use only classic facts and we always point to a place in \cite{prf} where they can be found.
From algorithmic information theory we use only the original ideas and notions from Kolmogorov \cite{Kol65,Kol68}
modified to a time-bounded version of Levin \cite{Lev84}. Standard notions from computational complexity
(as are classes P, NP, one-way permutations or pseudo-random generators) can be found in any textbook.

\section{Time optimality} 
\label{1}

The first thing that comes to mind is perhaps to compare two proof search algorithms by the time
they use. This is analogous to the fact that
in the optimality problem we compare two proof systems by the growth rate
of their lengths-of-proofs functions, i.e. by the non-deterministic time. 
For a deterministic algorithm $A$ that stops on all inputs
we denote by $time_A(w)$ the time $A$ needs to stop on input $w$.

We shall assume that proof search algorithms stop on every input, not just on inputs from TAUT. Namely,
if $A$ is an algorithm that stops on TAUT but maybe not everywhere outside TAUT, define new algorithm
$A'$ that in even steps computes as
$A$, and stops if $A$ does, and in odd steps performs an exhaustive search for a falsifying assignment
and stops if it finds one before $A$ stopped. The time complexity of $A$ and $A'$
on inputs from TAUT are proportional and $A'$ stops everywhere.

Note that in the following definition
the two proof search algorithms do not necessarily use the same proof system.

\begin{definition}
For two proof search algorithms $(A,P)$ and $(B,Q)$ define
$$
(A,P) \geq_t (B,Q)
$$
iff $(A,P)$ has at most polynomial slow-down over $(B,Q)$:
$$
time_A(\tau) \ \le \ time_{B}(\tau)^{O(1)}\ .
$$
for all $\tau \in \mbox{TAUT}$
(the constant implicit in $O(1)$ depends on the pair of the algorithms).
\end{definition}

\begin{lemma} \label{4.2.21b}
For any fixed proof system $P$ there is algorithm $A$ such that $(A,P)$ is 
$\geq_t$-maximal among all $(B,P)$, i.e.
it is {\bf time-optimal} among all
$(B,P)$. 
\end{lemma}

\prf

This is proved analogously as the existence of a universal NP search algorithm
(cf. Levin \cite{Lev73}): given input $\tau$, A tries 
for $i = 1, 2, \dots$
lexicographically first $i$ algorithms for $i$ steps until it finds a $P$-proof of $\tau$. 

We may assume w.l.o.g. that the size of the $i$-th algorithm is $O(\log i)$ and that $A$ simulates 
its $t$ steps in time polynomial in $t + \log i$.
Hence for a fixed $B$ that is $j$-th in the ordering then
$$
time_A(\tau) \le time_B(\tau)^{O(1)}\ .
$$

\qed

\noindent
{\bf Notation:}  {\em Let $(A_P,P)$ denote the proof search algorithm described in the above proof. 
Hence $(A_P,P)$ is time-optimal among all $(B,P)$.}

\bigskip

The optimality problem (both its versions for $\geq$ and $\geq_p$) relates to a number of 
questions in a surprisingly varied areas and there are quite a few relevant statements
known cf. \cite[Chpt.21]{prf}). We shall recall just one statement that we will use in the second proof
of Theorem \ref{6.1.21a}.

\begin{theorem} [{K. and Pudl\' ak \cite[Thm.2.1]{KraPud89a}}] \label{6.1.21b}
{\ }

A p-optimal proof system exists iff there is a deterministic algorithm $M$ computing the characteristic
function $\chi_{TAUT}$ of TAUT such that for any other deterministic algorithm $M'$ computing 
$\chi_{TAUT}$ it holds that:
$$
time_M(\tau) \le time_{M'}(\tau)^{O(1)}\ ,\ \mbox{ for all } \tau \in \mbox{TAUT}\ .
$$
\end{theorem}

Now we shall relate the existence of p-optimal proof systems and time-optimal proof search
algorithms. We give two proofs as they illustrate different facets of the statement.

\begin{theorem} \label{6.1.21a}
Let $P$ be any proof system containing resolution R and having the property that for some 
$c \geq 1$, for every $\tau$ and every $\tau'$ obtained from 
$\tau$ by substituting constants for some atoms it holds $s_P(\tau') \le s_P(\tau)^c$.

Then $P$ is p-optimal iff $(A_P,P)$ is time-optimal among all proof search algorithms $(B,Q)$.

In particular, a p-optimal proof system exists iff
a time-optimal proof search algorithm (i.e. $\geq_t$-maximal) exists.

\end{theorem}

\aprf

The only-if-direction is obvious, using Lemma \ref{4.2.21b}.
For the non-trivial if-direction of the theorem we use the fact that for any $Q$ there is a 
{\em p-time construable sequence} of tautologies
$$
\langle Ref_Q\rangle_n\ ,\ n \geq 1
$$
such that $n \le |\langle Ref_Q\rangle_n|$ and
if $P$-proofs of these formulas are p-time computable then $P$ p-simulates $Q$. 
These formulas formalize the soundness\footnote{They have bits for size $\le n$ $Q$-proof $x$, formula $y$ and truth assignment
$z$ and say that if $x$ is a $Q$-proof of $y$ then $z$ satisfies $y$, cf. \cite{prf}.}
 of $Q$ and their exact definition is not important here.
Their relation to (p-)simulations 
is a classic fact of proof complexity going back to Cook \cite{Coo75}; 
see \cite[Secs.19.2 or 21.1]{prf} for this background.

Define a proof system $Q'$ in which $1^{(n)}$ is a proof of $\langle Ref_Q\rangle_n$ and any other $w$
is interpreted as a resolution proof. Further take algorithm $B$ which upon receiving
$\tau$ first looks whether $\tau = \langle Ref_Q\rangle_n$ for some $n$ (a priori $\le |\tau|$)
in which case it produces $1^{(n)}$, and otherwise it uses some fixed resolution searching algorithm to find a proof.

Because $(A_P,P)$ is supposed to be time optimal, $A_P(\langle Ref_Q\rangle_n)$ has to
compute in p-time a $P$-proof of $\langle Ref_Q\rangle_n$. But by the stated properties of these formulas
$P \geq_p Q$ follows.  

\bigskip

\bprf

We now give a second, alternative proof for the last statement of the theorem, using Theorem \ref{6.1.21b}.
For a proof search algorithm $(A,P)$ define algorithm $M_{(A,P)}$ computing $\chi_{TAUT}$
as follows: On input $\tau$ it computes $A(\tau)$ and checks that
$P(A(\tau)) = \tau$. If so, it outputs $1$, otherwise it outputs $0$.

On the other hand, if $M$ computes $\chi_{TAUT}$ define proof system $P_M$ by
$$
P_M(w) = \tau\ \mbox{ iff }\ 
\mbox{ {\em ($w$ is the computation of $M$ on $\tau$ and it outputs $1$)} }\ 
$$
and algorithm $A_M$: On input $\tau$ output the computation of $M$ and $\tau$.

It is easy to verify that:

\begin{itemize}

\item if $(A,P)$ is a time-optimal proof search algorithm then $M_{(A,P)}$ is a deterministic
algorithm computing $\chi_{TAUT}$ having the time-optimality property from Theorem \ref{6.1.21b},
and

\item if $M$ is a deterministic
algorithm computing $\chi_{TAUT}$ having the time-optimality property from Theorem \ref{6.1.21b}
then $(A_M, P_M)$ is time-optimal proof search algorithm.

\end{itemize}
\noindent
Theorem \ref{6.1.21b} then implies the statement.

\qed

If $P \geq_p Q$ then in any reasonable quasi-ordering of proof search algorithms
$(A_p,P)$ will be at least as strong as $(A_Q,Q)$. For the opposite direction (the if-direction)
of the theorem we utilized the fact that $(A_P,P)$ is required to find in p-time proofs 
of {\em simple sequences} of formulas as are $\langle Ref_Q\rangle_n$.
A simple sequence of formulas appears also in the following situation. 
Take any proof search algorithm $(A,\mbox{R})$ searching for
resolution proofs. Take a sequence of tautologies that are computed by a p-time function from $1^{(n)}$ 
and that are hard for R 
but easy for {\bf Extended resolution ER} and, moreover, their ER-proofs can be computed
from $1^{(n)}$ in p-time by some function $f$. Examples of such formulas are formulas $PHP_n$ formalizing the
pigeonhole principle, cf. Haken \cite{Hak85} and Cook and Reckhow \cite{CooRec} (or see \cite{prf}).  
Now define a proof search algorithm $(B,\mbox{ER})$ that on input $\tau$ computes as follows:

\begin{enumerate}

\item $B$ checks if $\tau = PHP_n$ for some $n \geq 1$ (this is p-time because
it needs to consider only $n \le |\tau|$).

\item If yes, i.e. $\tau = PHP_n$, then $B$ outputs $f(1^{(n)})$.

\item Otherwise $B$ outputs $A(\tau)$.
\end{enumerate}
Then $(B,ER) >_t (A,R)$ but intuitively it does not seem quite right to claim
that $(B,ER)$ is a better algorithm than $(A,R)$; $B$ does not do anything extra except that it
remembers one type of simple formulas. One would like to 
\begin{itemize}

\item [(*)] compare $A$ and $B$ on inputs $\tau$
where {\em they actually do something non-trivial}.

\end{itemize}
In \cite[Sec.21.5]{prf} we proposed a definition of a quasi-ordering of proof search algorithms
by time as is $\geq_t$ but measured only on TAUT from which we are allowed to take out
a simple (in particular, a p-time construable) sequence of tautologies.
Subsequently in \cite{Kra-mfo} a stronger variant of 
that (avoiding all such sequences) was proposed. This could, in principle, allow for the situation that there is 
an optimal proof search algorithm without having a p-optimal proof system, and thus separate the two questions.
However, the resulting quasi-orderings are unintuitive and it is not clear whether they actually help
to avoid the if-direction of Theorem \ref{6.1.21a}.

A more fundamental issue, related to (*) above, is that the decision not to count (or not count)
$\tau = PHP_n$ when comparing 
two proof search algorithms is not based only on the individual tautology $\tau$ but depends on the fact that
it is one of an infinite series of tautologies defined in a particular uniform way.

These considerations are, of course, quite informal but lead us to formal notions discussed in the next section.

\section{Information optimality} \label{2}

We shall assume that every $e \in \uu$ is also a code of a unique Turing machine and we shall consider
a universal Turing machine $U$ with three inputs $e,u,1^{(t)}$ that simulates machine $e$ on input
$u$ for at most $t$ steps, stops with the same output if $e$ stops in $\le t$ steps, and otherwise outputs
$0$. We shall assume that $U$ runs in polynomial time.

Using this set-up 
recall {\bf the time-bounded Kolmogorov complexity} of a string $w \in \uu$ as defined by Levin \cite{Lev73}:
$$
Kt(w|u)\ :=\ \min \{(|e| + \lceil \log t\rceil)\ |\  U(e,u,1^{(t)}) = w \}
$$
(we use $\lceil \log t\rceil$ instead of $\log t$ as we want integer values)
and
$$
Kt(w)\ := Kt(w|0)\ .
$$
Intuitively, smaller $Kt(w)$ is simpler $w$ is, in the sense that it can be compressed to a shorter string without loosing
information.

Note that we have trivial estimates to $Kt(w|u)$ and $Kt(w)$ in terms of the size $|w|$:
\begin{equation} \label{11.2.21b}
\log(|w|) \le Kt(w|u) \le Kt(w) \le |w| + \log(|w|) + O(1)\ .
\end{equation}
The left inequality holds as need time $|w|$ to write $w$, the middle one is trivial and the right
inequality follows from considering a machine that has $w$ hardwired into its program. 

We would like to have inequality $Kt(w) \le Kt(w|u) + Kt(u)$ that is intuitively justified by composing
machine $e_1$ computing $u$ with machine $e_2$ computing $w$ from $u$. However, as pointed out in Kolmogorov \cite{Kol68},
the code of the composed machine (and, in general, of the pair $(e_1,e_2)$) does not have length $|e_1| + |e_2|$
but rather can be defined of length  $|e_1| + |e_2| + O(\log(|e_1|) + \log(|e_2|))$. Hence we get a slightly
worse inequality:
\begin{equation} \label{21.2.21b}
Kt(w) \le Kt(w|u) + Kt(u) + O(\log(Kt(w|u)) + \log(Kt(u)))  
\end{equation}
and similarly
\begin{equation} \label{11.2.21a}
Kt(w|u) \le Kt(w|v) + Kt(v|u) + O(\log(Kt(w|v)) + \log(Kt(v|u)))\ .  
\end{equation}

\bigskip
\noindent
We use $Kt$ to define a new measure of complexity of proofs.

\begin{definition}
Let $P$ be a proof system. For any $\tau \in TAUT$ define
$$
i_P(\tau)\ :=\ 
\min \{Kt(w|\tau)\ |\ w \in \bits^* \wedge P(w) = \tau\} \ . 
$$
We shall call $i_P$ {\bf the information efficiency function}.
\end{definition}

The function measures the minimal amount of information any $P$-proof of $\tau$ has to contain,
{\em knowing what $\tau$ is}. The next statement shows that stronger proof system do not 
require much more information.

\begin{lemma} \label{21.2.21a}
For any $P, Q$, $P \geq_p Q$ implies
$i_P(\tau) \le O(i_Q(\tau))$.
\end{lemma}

\prf

Let $f$ be a p-simulation of $Q$ by $P$. Take $w$ that is a $Q$-proof of $\tau$
with $Kt(w|\tau) = i_Q(\tau)$.

Using (\ref{11.2.21a}) we can estimate $i_P(\tau) \le Kt(f(w)|\tau)$ from above by the sum 
$Kt(f(w)|w) + Kt(w|\tau) = Kt(f(w)|w) + i_Q(\tau)$ plus some log-small terms. 
But $Kt(f(w)|w) \le O(\log |w|) + O(1)$ (the $O(1)$ is for the machine computing $f$
and the computation runs in time polynomial in $|w|$)
which is also bounded by $O(i_Q(\tau))$ by (\ref{11.2.21b}).

\qed

The next two statements relate the information measure fairly precisely to time in proof search.

\begin{lemma} \label{4.2.21a}
Let $(A,P)$ be any proof search algorithm. Then for all $\tau \in TAUT$:
$$
i_P(\tau) \le Kt(A(\tau)|\tau) \le |A| + \log (time_A(\tau))\ .
$$
In particular, $time_A(\tau) \geq \Omega(2^{i_P(\tau)})$.

\end{lemma}

\prf

The first inequality is obvious, the second follows from 
the definition as $A(\tau) = U(A, \tau, 1^{(t)})$, where 
$t = time_A(\tau)$.

\qed

This statement is complemented by the next one essentially saying that {\em easy proofs are easy to
find\footnote{In this sense it establishes automatizability of all proof systems w.r.t. information
efficiency as opposed to the original automatizability relating to 
lengths-of-proofs, cf. Sec. \ref{3}  or \cite[Sec.17.3]{prf}.}}.

\begin{lemma} [i-automatizability]  \label{4.2.21c}
{\ }

For every proof system $P$ there is an algorithm $B$ such that for all $\tau \in TAUT$:
$$
Kt(B(\tau)|\tau) = i_P(\tau)
$$
and 
$$
time_B(\tau) \le 2^{O(i_P(\tau))}\ .
$$
\end{lemma}

\prf

For $i = 1,2, \dots$ algorithm $B$ (using the universal machine $U$) does the following:

\begin{itemize}

\item {\em In the lexico-graphic order tries all pairs $(e,t)$ such that $|e| + \lceil \log t \rceil = i$
and checks whether $U(e, \tau, 1^{(t)})$ is a $P$-proof of $\tau$.
If so, it outputs the proof and $B$ stops.}

\end{itemize}
There are $\le 2^{2i}$ such pairs $(e,t)$ to consider, computing 
$U(e, \tau, 1^{(t)})$ takes time $poly(|e|, t) \le 2^{O(i)}$ and checking
whether $P(U(e, \tau, 1^{(t)})) = \tau$ takes time\newline
$poly(|U(e, \tau, 1^{(t)})|) \le 2^{O(i)}$.
The procedure takes for one $i$ overall time $2^{O(i)}$ and because $B$ succeeds in the round for
$i = i_P(\tau)$, the overall time $B$ takes is $\le 2^{O(i_P(\tau))}$.

\qed

\noindent 
{\bf Notation:} 
{\em Let us denote the algorithm described in the proof by $B_P$. }

\bigskip

In fact, the argument in the proof of Lemma \ref{4.2.21c}
is another version of the universal search as the next statement shows.

\begin{corollary}
Let $P$ be any proof system and let $A_P$ and $B_P$ be the two algorithms defined earlier.
Then
$$
(A_P,P) \geq_t (B_P,P) \geq_t (A_P,P)\ . 
$$
\end{corollary}

\prf

The first inequality follows from Lemma \ref{4.2.21b}, and the second from Lemmas 
\ref{4.2.21a} and \ref{4.2.21c}.

\qed

Because the algorithm $B_P$ achieves the optimal information efficiency 
it seems natural to define
a quasi ordering of proof systems based on comparing their information-efficiency
functions.

\begin{definition}
For two proof systems $P$ and $Q$ define:
$$
P \geq_i Q\ \mbox{ iff }\ 
i_P(\tau) \le O(i_Q(\tau))
$$
for all $\tau \in TAUT$.

\end{definition}

This is a quasi ordering of proof systems that is, by Lemma \ref{21.2.21a}, 
coarser that $\geq_p$ but, presumably, different than both $\geq_p$ and $\geq$.
But as far as optimality goes it does not allow for a new notion.

\begin{theorem} \label{14.2.21a}
Let $P$ be any proof system containing resolution R and having the property that for some 
$c \geq 1$, for every $\tau$ and every $\tau'$ obtained from 
$\tau$ by substituting constants for some atoms it holds $s_P(\tau') \le s_P(\tau)^c$.

Then $P$ is information-optimal (i.e. $\geq_i$-maximal) iff it is  p-optimal.
\end{theorem}

\prf

Let $P$ be a p-optimal proof system and let $Q$ be any proof system. Assume $f$ is a p-simulation of
$Q$ by $P$.

Let $\tau \in TAUT$ and
assume $Kt(w|\tau) = i_Q(\tau)$ for some $Q$-proof $w$ of $\tau$. Then $f(w)$ is a $P$-proof
of $\tau$ and $Kt(f(w)|w) \le O(1) + O(\log |w|)$.
But $|w| \le 2^{i_Q(\tau)}$, so
$Kt(f(w)|w) \le O(i_Q(\tau))$ and $Kt(f(w)|\tau) \le O(i_Q(\tau))$ follows
by (\ref{11.2.21a}). Hence
$$
i_P(\tau) \le O(i_Q(\tau))\ ,\ \mbox{ all }\ \tau \in TAUT\ .
$$

\bigskip

For the only-if-direction assume that $P$ is an information-optimal proof system and $Q$ is an arbitrary 
proof system. Take the sequence $\langle Ref_Q\rangle_n$, $n \geq 1$, as in the first proof
of Theorem \ref{6.1.21a}, and interpret strings $1^{(n)}$ as proofs of these formulas in some proof system
$Q'$. We see that
$$
i_{Q'}(\langle Ref_Q\rangle_n) \le O(\log n)\ .
$$
By the information optimality of $P$ also 
$$
i_{P}(\langle Ref_Q\rangle_n) \le O(\log n)
$$
which, by Lemma \ref{4.2.21c}, 
means that the algorithm $B_P$ finds $P$-proofs of formulas $\langle Ref_Q\rangle_n$
in time $n^{O(1)}$. This implies, as in the first proof of Theorem \ref{6.1.21a}, that
$P \geq_p Q$.

\qed

Theorem \ref{14.2.21a} implies that the information measure approach does not lead to
a separation of the proof search problem from the optimality problem either. 

\section{Information vs. size} \label{3}

A natural question is whether the information-efficiency function may give, at least in principle, better
time lower bounds for proof search algorithms than the length-of-proof function. By Lemma \ref{4.2.21a}
information gives super-polynomially better time lower bound than size if
$i_P(\tau)$ cannot be in general bounded above by $O(\log s_P(\tau))$.

Recall a notion introduced by Bonet, Pitassi and Raz \cite{BPR00}: a
proof system $P$ is {\bf automatizable} iff there is a proof search algorithm $(A,P)$ such that for
all $\tau \in \mbox{TAUT}$:
$$
time_A(\tau) \le s_P(\tau)^{O(1)}\ .
$$
Considering that there are no known non-trivial complete automatizable proof systems 
this author saw as the only use of the notion that it gives a nice meaning to
the failure of feasible interpolation, cf. \cite[Sec.17.3]{prf}. But now it is exactly what we need  
to characterize the separation of size from information; using Lemma \ref{4.2.21a} 
the following statement is obvious.

\begin{theorem} \label{13.2.21a}
A proof system $P$ is non-automatizable iff 
there is an infinite set $X$ of tautologies $\tau$ of unbounded size such that
\begin{equation} \label{6.4.21a} 
i_P(\tau) \geq \omega(\log s_P(\tau))
\end{equation}
on $X$. 
\end{theorem}

To illustrate what type of formulas witness the separation of size from information we
shall paraphrase the construction from \cite{KP-crypto}; there it was done for $P = \mbox{ER}$ and $h := \mbox{RSA}$.

Let $h : \uu \rightarrow \uu$ be a p-time permutation of each $\nn$, i.e. it is a length-preserving and injective
function, and let $h_n$ be the restriction of $h$ to $\nn$. 
For any $b \in \nn$ define formula
$$
\mu_b\ :=\ [h_n(x) = b \rightarrow B(x)=B(h^{(-1)}(b))]\ 
$$
where $B(x)$ is a hard-bit of permutation $h$; 
the statement $h_n(x) = b$ is expressed by a p-size circuit (if $P$ allows them), or using
auxiliari variables whose values are uniquely determined by values of $x_1, \dots, x_n$.
Note that $|\mu_b| \le n^{O(1)}$.

\begin{lemma}
Let $P$ be any proof system containing resolution R and having the property that for some 
$c \geq 1$, for every $\tau$ and every $\tau'$ obtained from 
$\tau$ by substituting constants for some atoms it holds $s_P(\tau') \le s_P(\tau)^c$.

Assume that $P$ proves by p-size proofs tautologies expressing
that $h_n$ are injective:
$$
h_n(x) = h_n(y) \rightarrow \bigwedge_{i \le n} x_i \equiv y_i\ .
$$
Assume that $h$ is a one-way permutation and $B$ is its hard bit predicate\footnote{See
\cite{Pap-book} for definitions of these notions.}. 

Then there are $P$-proofs $\pi_b$ of formulas
$\mu_b$ such that:
\begin{enumerate}

\item $|\pi_b| \le n^{O(1)}$, i.e. $s_P(\mu_b)\ \le\ n^{O(1)}$, 

\item for a random $b \in \nn$, with a probability going to $1$ as $n \rightarrow \infty$, 
it holds that
\begin{equation} \label{14.2.21b}
i_P(\mu_b) \geq \omega(\log n)\ .
\end{equation}
If $h$ is secure even against algorithms running in time $2^{n^\epsilon}$, for some $\epsilon > 0$,
then the right-hand term in (\ref{14.2.21b}) can be improved to $n^{\Omega(1)}$.

\end{enumerate}

\end{lemma}

\prf

Define the wanted $P$-proof $\pi_b$ as follows. Pick $a \in \nn$ such that $h(a) = b$ and prove
in $P$ in p-size, using the injectivity of $h_n$, that
$$
h_n(x) = b \rightarrow x = a\ .
$$
From this implication we can derive $\mu_b$ using implication
$$
x = a \rightarrow B(x) = B(a)
$$
that has p-size resolution proofs.
This proves the first statement.

The second statement follows from the hypothesis that $h$ is one-way:
we can try the algorithm $B_P$ from Section \ref{2} on formulas
$$
[h_n(x) = b \rightarrow B(x)= c]
$$
for $c = 0, 1$ and compute in this way the hard bit in p-time. But that is impossible if $B$ is indeed 
a hard bit of $h$.

\qed

Related formulas can be defined as follows.
Let $\varphi_n(x)$, $n \geq 1$ and $x = (x_1, \dots, x_n)$,
be a sequence of formulas that have p-size $|\varphi_n| \le n^{O(1)}$ but that do not have p-size $P$-proofs:
$$
s_P(\varphi_n) \geq n^{\omega(1)}\ ,\ n \geq 1\ .
$$
For some proof systems we have such formulas unconditionally, for those which are not p-optimal we
can take formulas $\langle Ref_Q\rangle_n$ used earlier, for some $Q >_p P$.

For any $b \in \nn$ define formula
$$
\eta_b(x)\ :=\ [h_n(x) = b \rightarrow \varphi_n(x)]\ .
$$
Note that $|\eta_n| \le n^{O(1)}$. Analogously with the proof of the lemma, the formulas have p-size $P$-proofs
$\pi_b$ and these particular proofs satisfy $Kt(\pi_b | \eta_b) \geq \omega(\log n)$. It would be interesting 
if for some $P$ it would hold that any short proof of $\eta_b$ must contain some non-trivial information 
about $h^{(-1)}(b)$.

\section{Information alone} \label{5.4.21a}

A separation of size from information in the sense of 
(\ref{6.4.21a})
implies that no p-time algorithm finds, given $\tau$ and $s_P(\tau)$ in unary,
a p-time recognizable (by $P$) object (a $P$-proof), and hence it implies that $\mbox{P} \neq \mbox{NP}$. 
In fact, a number of proof systems are known to be non-automatizable assuming various conjectures from
complexity theory (cf. \cite[Sec.17.3]{prf}). We mention just resolution R and its non-automatizability proved under
the weakest possible hypothesis that $\mbox{P} \neq \mbox{NP}$ by Atserias and M\"{u}ller \cite{AtsMul20}; references for 
earlier work and other examples can be found there or in \cite[Sec.17.3]{prf}.

Proofs of non-automatizability depend on a p-time reduction of some hard set $Y$ (NP-complete in \cite{AtsMul20} or
hard bit of RSA in \cite{KP-crypto} or similar, cf. \cite[Sec.17.3]{prf}) to a set of formulas with p-size proofs
that maps the complement $\uu \setminus Y$ to formulas with only long (or none) proofs.
These arguments do not yield lower bounds for $i_P(\tau)$ for individual formulas but only speak about the asymptotic 
behavior of an automatizing algorithm.

\bigskip

We are interested in the question whether one can establish a lower bound for $i_P(\tau)$ by considering formulas
individually, not as members of an infinite set or sequence. This would be
in a way analogous to lengths-of-proofs lower bounds (e.g. for $PHP_n$ in R in \cite{Hak85}) which 
work with individual formulas. 

A super-polynomial lower bound for $s_P(\tau)$ is used primarily for
three purposes:
\begin{enumerate}

\item {\em It implies that no $Q \le P$ is p-bounded, an instance of $\mbox{NP} \neq \mbox{coNP}$, and if true
for all $P$ then indeed $\mbox{NP} \neq \mbox{coNP}$ follows.}

\item {\em It implies super-polynomial time lower bounds for a class of SAT algorithms $S$ that are
simulated by $P$: $P \geq P_S$ ($P_S$ defined in the Introduction). Currently known lengths-of-proofs lower
bounds imply time lower bounds for large classes of SAT algorithms.}

\item {\em It implies independence results from a first-order theory attached to $P$ and, in particular, 
that $\mbox{P} \neq \mbox{NP}$ is consistent with the theory (see \cite[Sec.86]{prf}}).

\end{enumerate}
But having a super-logarithmic lower bound for $i_P(\tau)$ is just as good.
Items 2. and 3. hold literally: in the former this is by Lemma \ref{4.2.21a} and for the latter
this holds because propositional translations of first-order proofs are performed by p-time algorithms
(cf. \cite[Part 2]{prf}). In item 1 one has to compromise on weakening
$\mbox{NP} \neq \mbox{coNP}$ to $\mbox{P} \neq \mbox{NP}$.

This motivates the following problem that seems to us to be quite fundamental.

\begin{problem} \label{9.4.21a}
Establish {\em unconditional} super-logarithmic lower bound
$$
i_P(\tau) \geq \omega(\log |\tau|)
$$
for $\tau$ from a set $X \subseteq \mbox{TAUT}$ of tautologies of unbounded size, for a proof system $P$ for which no
super-polynomial lowers bounds for the length-of-proof function $s_P$ are known.
\end{problem}
As a step towards solving the problem it would be interesting to have such unconditional lower bounds 
at least for $P$ for which super-polynomial
lower bounds for $s_P$ are known, but not for formulas from $X$.

Note the emphasis on the requirement that the lower bound is unconditional. Allowing some unproven computational complexity
hypotheses the problem becomes easy. For example, if it were that $i_P(\tau ) \le O(\log |\tau|)$ for all $\tau$
then the algorithm $B_P$ form Section \ref{2} runs in p-time and hence $\mbox{P} = \mbox{NP}$. Or you may take
any pseudo-random number generator $g : \nn \rightarrow \bits^{n+1}$ and for $b \in \bits^{n+1}$
take a formula\footnote{See Subsection \ref{9.4.21b}.} $\tau_b$ expressing that $b \notin Rng(g)$.
Then $i_P(\tau_b)$ cannot be bounded by $O(\log |\tau_b|)$ as otherwise $B_P$ would break
the generator in p-time.

\bigskip

In what follows we shall discuss the existence of formulas $\tau$ whose length we shall denote
$m$. The formulas will not be a priori members of some infinite series but are considered {\em individually}.
This means that questions and statements about them do depend just on them and not on asymptotic properties
of some ambient sequence. But we still wish to use
the handy $O$-, $\Omega$- and $\omega$- notations  and in doing so
we imagine what happens in each particular construction or statement as $m \rightarrow \infty$.

For the sake of the following discussion let us call a size $m$ formula {\bf simple} if $Kt(\tau) = O(\log m)$
and {\bf complex} otherwise, and we apply similar qualifications to its proofs $\pi$ but still
relative to parameter $m$ (i.e. not relative to $|\pi|$). 

For example, for the truth-table proof system TT, any tautology $\tau$ in $m^{\Omega(1)}$ variables, simple or complex, 
will have only a complex truth-table proof $\pi$: its size is exponential in $m^{\Omega(1)}$
and (\ref{11.2.21b}) implies that $Kt(\pi) \geq i_{TT}(\tau) \geq m^{\Omega(1)}$ as well.

To solve Problem \ref{9.4.21a} we want a class $X \subseteq \mbox{TAUT}$
of formulas $\tau$, $|\tau| = m \rightarrow \infty$, such that
$$
i_P(\tau) \geq \omega(\log m)\ .
$$
The following lemma formulates two simple conditions on $X$, one necessary and one sufficient.

\begin{lemma}
Let $X \subseteq \mbox{TAUT}$ be a set of formulas of unbounded size.

\begin{enumerate}

\item (a necessary condition) 

For $X$ to solve Problem \ref{9.4.21a} it is necessary that all $P$-proofs
$\pi$ of all $\tau \in X$ are complex:
$$
Kt(\pi) \geq \omega(\log m)\ .
$$

\item (a sufficient condition)

If $X$ satisfies item 1 then a sufficient condition for it to solve the problem
is that all $\tau \in X$ are simple:
$$
Kt(\tau) \le O(\log m)\ .
$$

\end{enumerate}

\end{lemma}

\prf

For item 1 note that
by (\ref{11.2.21b}) we have
$i_P(\tau) \le Kt(\pi|\tau) \le Kt(\pi)$.
For item 2 we have by (\ref{21.2.21b}): 
$$
Kt(\pi) \le Kt(\pi|\tau) + Kt(\tau) + O(\log Kt(\pi|\tau)) + O(\log Kt(\tau))\ .
$$
By (\ref{11.2.21b}) we may 
estimate the last term by $O(\log m)$ for any $\tau$, and by the hypothesis $Kt(\tau) \le O(\log m)$ as well. 
Hence we can rewrite the inequality as
\begin{equation} \label{9.4.21d}
Kt(\pi) - Kt(\pi|\tau) \le O(\log m) + O(\log Kt(\pi|\tau))\ .
\end{equation}
Now distinguish two cases. Either $\pi \le m^{O(1)}$ or 
$\pi \geq m^{\omega(1)}$. In the latter case we are done as
$i_P(\tau)$ is lower bounded by $\log s_P(\tau)$. In the former case
we can estimate the last term in (\ref{9.4.21d}) by $O(\log m)$ and hence get
\begin{equation} \label{9.4.21e}
Kt(\pi) - Kt(\pi|\tau) \le O(\log m)\ .
\end{equation}
This implies what we need because, by item 1, $Kt(\pi) \geq \omega(\log m)$.

\qed

Note that condition 1 in the lemma is not sufficient. To see this take
$\tau$ of the form $\rho \vee \neg \rho$, where $\rho$ is random a hence of high $Kt$-complexity. But
$\tau$ is a proof of itself in a suitable Frege system (or even in R if $\rho$ is just a clause
and $\neg \rho$ is the set of singleton clauses consisting of negations of literals in $\rho$)
and $Kt(\tau|\tau) = \log(|\tau|) + O(1)$ is small.

\bigskip

When $\tau$ are complex then 
the necessary condition holds automatically: given a $P$-proof $\pi$ of $\tau$, either
$|\pi| \geq m^{\omega(1)}$ and hence $\omega(\log m)$ lower bounds $Kt(\pi|\tau)$ by 
(\ref{11.2.21b}), or $|\pi| \le m^{O(1)}$. In the latter case, because $P(\pi) = \tau$
and using  (\ref{21.2.21b}):
$$
Kt(\tau) \le Kt(\tau|\pi) + Kt(\pi) + O(\log Kt(\tau|\pi) + \log Kt(\pi))
$$
which yields
$$
\omega(\log m) \le O(1) + O(\log |\pi|) + Kt(\pi) + O(\log (O(1) + O(\log |\pi|)) + \log Kt(\pi))\ .
$$
Estimating $\log |\pi| \le O(\log m)$ we derive:
$$
\omega(\log m) \le Kt(\pi)\ .
$$
On the other hand, the computation in the proof of item 2 does not yield anything for complex formulas. But the quantity
being estimated from above in (\ref{9.4.21e}) still makes sense and if (\ref{9.4.21e})
holds for an $X$ (satisfying item 1) then $X$ solves the problem.

In fact, this quantity has been isolated already by
Kolmogorov \cite{Kol65,Kol68}; following 
him define (the $Kt$-version of)
{\bf information that $u$ conveys about $w$} as
$$
It(u : w) \ :=\ Kt(w) - Kt(w|u)\ .
$$
Hence what we want is $\tau$, simple or complex, having only complex proofs 
such that for any proof $\pi$ it holds that:
$$
It(\tau : \pi) \ \mbox{ {\em is small}. }
$$
In words: $\tau$ knows very little about its proofs.

\bigskip

Many formulas that appear in various contexts of proof complexity 
(as formulas $PHP_n$ or $\langle Ref_Q\rangle_n$ we encountered earlier),
occur as members in a uniformly constructed sequence $\{\tau_n\}_n$.
The sequence is often p-time construable from $1^{(n)}$ or, in fact, have even stricter levels 
of uniformity (cf. \cite[Sec.19.1]{prf}).
When such formulas have short proofs $\pi_n$ in some proof system $P$ it is often the case that the proofs are
also uniformly constructed from $1^{(n)}$. But that forces $Kt(\pi_n)$ to be $O(\log n)$.

Hence if we wanted to use for $X$ some uniform formulas they ought to be expected to have only long $P$-proofs
(but we may not be able to prove that). Leaving the reflection principles aside, two examples that
come to mind are
\begin{itemize}

\item $AC^0[p]$-Frege systems and the $PHP_n$ formulas, cf. \cite[Sec.10.1 and Problem 15.6.1]{prf}.

(No super-polynomial size lower bounds are known for this proof system, cf. \cite[Problem 15.6.1]{prf}.)

\item $AC^0$-Frege systems and the $WPHP_n$ formulas (expressing weak PHP).

(Lower bounds for $AC^0$-Frege systems are known but not for formulas $WPHP_n$ expressing a form
of the weak PHP, cf. \cite[Problem 15.3.2]{prf}.)
\end{itemize}
For stronger systems the only candidates for hard formulas\footnote{We leave reflection principles
aside here.} which are supported by some
theory are {\bf $\tau$-formulas}, called also {\bf proof complexity generators}. 
These formulas are described in Subsection \ref{9.4.21b}. For some generators, as those defined in
\cite[Secs. 29.4-5]{k2} these formulas are expected to be all complex in the sense of $Kt$ complexity.

But for the $\tau$-formulas based on the truth-table function $\tru_{s,k}$ there are uniform
examples possibly hard for ER (Extended resolution). The formulas express that a size $2^k$
string is not the truth-table of a Boolean function computed by a size $\le s$ circuit.
Truth tables of SAT (in fact, of any language in the class E) 
are constructible in p-time (i.e. in time polynomial in $2^k$) and 
there is a theory (cf. \cite[Sec.5]{Kra-dual})
supporting the conjecture that the corresponding $\tau$-formulas are 
hard for ER. 

The theory of proof complexity generators is now fairly extensive and it is not feasible to repeat
its key points here.
More information is in Subsection \ref{9.4.21b} and in references given there.

\section{Proof complexity remarks} \label{4}

In this section we remark on several topics in proof complexity that seem 
to be related to the information measure. It may be worthwhile to explore if there
are some deeper connections.
The section aims primarily at proof complexity readers but we 
give references to relevant places in \cite{prf} to aid non-specialists.

\subsection{Proof complexity generators} \label{9.4.21b}

A fairly succinct exposition of the theory of proof complexity generators
can be found in \cite[Secs.19.4 and 19.6]{prf} or in older
\cite[Chpts.29 and 30]{k2}.
The theory investigates, in particular, functions $g$ extending $n$ bit strings to $m$-bit strings, $m = m(n) > n$, 
computable in p-time, and such that formulas
$\tau(g)_b$, expressing for $b \in \bits^m$ that $b \notin Rng(g)$, ought to be hard to prove in a given proof system. 
In particular, function $g$ is defined to be {\bf hard for $P$}
iff for any $c\geq 1$ only finitely many formulas $\tau(g)_b$ have a $P$-proof of
size $\le |\tau(g)_b|^c$.

Function $g$ can be thought of
as a decompression algorithm and for $w \in Rng(g)$ we have $Kt(w) \le n + O(\log n) + O(1)$ which is $<< m$ if, for example,
$3n \le m$. 
Note that for $w \in \mm$, $3n \le m$, the condition $Kt(w) \geq m/2$ implies 
that $Kt(w) > n + O(\log n) + O(1)$
and hence also $w \notin Rng(g)$. The property
$Kt(w) \geq m/2$ cannot be expressed by a p-size tautology as the time involved in the computation of
the universal machine may be exponential in $m$.  
But for a fixed p-time $t(n)$ we can consider complexity $K^t$ by restricting the decompression to a universal Turing
machine $U^t$ on inputs $e,u$ (i.e. no time input) simulating $e$ on $u$ for time $t$. By padding (or restricting) 
all outputs in some canonical way to size $m = m(n)$ exactly,
and taking for the domain $n'$-bit strings with, say, $n':= n + \log n$ (the term $\log n$ swallowing the
description of a machine), we can think of $U^t$ as of a generator as well.

By the virtue of constructions of universal $U^t$ (for time $t$ machines) it is straightforward to show in theory PV that 
$Rng(g) \subseteq Rng(U^t)$ for any generator $g$ as above running in time $\le t(n)$.
Hence (the propositional translations of) this fact are shortly provable in ER, cf. \cite[Chpt.12]{prf}).
It follows that for any $P \geq ER$, if some $\tau$-formulas
resulting from $U^t$ have short proofs so do some formulas resulting from $g$. That is, if there is any 
$g$ computable in time $t$ and hard for $P$ then $U^t$ must be hard as well.
Putting it differently, proving tautologies\footnote{A referee pointed out that Pich and Santhanam \cite{PS}
considered similar tautologies expressing high Kolmogorov complexity based on the $KT$ complexity.} 
expressing $K^t(w) > m/2$ must be hard for $P$
\footnote{It is tempting to look for analogies of these tautologies with formulas occurring
in Chaitin's \cite{Chai74} incompleteness theorem. But the interpretation of the role of information
in incompleteness phenomenon is littered with mathematically unsupported or outright incorrect 
interpretations - see van Lambalgen \cite{Lam} for analysis of some - and we stay away from any
informal discussion of this topic.}.

\subsection{Implicit proof systems}

Implicit proof systems, introduced in \cite{Kra-implicit}, operate with proofs $\pi$ computed bit-by-bit
by a circuit (but that is not all). Proof $\pi$ may have size exponential in comparison with 
the size of the defining circuit. Hence its $Kt$-complexity may be close to the lower bound $\log |\pi|$
from (\ref{11.2.21b}).

For two proof systems $P, Q$ the {\bf implicit proof system}
$[P,Q]$ considers a proof of a tautology $\tau$ to be a pair $(\alpha,\beta)$, where
$\beta$ is a circuit whose truth-table is a $Q$-proof of $\tau$ and $\alpha$ is a $P$-proof
(of the propositional statement formalizing) that $\beta$ indeed computes a $Q$-proof.
Note that using circuits $\beta$ alone would not constitute a Cook-Reckhow proof system.
For the formal definition see \cite{Kra-implicit} or \cite[Sec.7.3]{prf}.

Implicit proof systems get incredibly strong very fast. For example, {\bf implicit resolution} $iR := [R,R]$
p-simulates ER and iER p-simulates quantified propositional system $G$, cf. \cite[Sec.7.3]{prf}.

\subsection{Proof systems with advice} \label{26.2.21a}

Recall from Cook and K. \cite{CooKra} that
a {\bf functional\footnote{Classical proof systems can be formulated either as functional - as we did at the beginning
of the paper - or as relational and these two formulations are essentially equivalent from proof complexity
point of view. This is no longer true for systems with advice, cf. \cite{CooKra}.}
proof system with $k(n)$ bits of advice} is a $P : \uu \rightarrow \uu$
whose range is exactly $\mbox{TAUT}$
and such that $P$ is computable in polynomial time  using $k(n)$
bits of advice on inputs (i.e. proofs) of length $n$.
Cook and K. \cite[Thm.6.6]{CooKra}  proved that
there exists a proof system with $1$ bit of advice that p-simulates
all classical Cook-Reckhow's proof systems. This suggests\footnote{To us it was suggested by
Igor C.~Oliveira.} to use the only-if direction of
Theorem \ref{6.1.21a} and to conclude that there is a {\em proof search algorithm with advice} $(A,P)$
which is $\geq_t$-better than all ordinary proof search algorithms of Definition \ref{27.2.21a}.
Here $P$ is the proof system with $1$ bit of advice from \cite[Thm.6.6]{CooKra} and $A$ is a 
{\em non-uniform} p-time algorithm, i.e. it uses p-size advice.

To see this note that the proof of the only-if direction in Theorem \ref{6.1.21a} appeals to Lemma \ref{4.2.21b}
that there is a time-optimal algorithm for any fixed proof system: in the universal search construction 
we need to check many - but only polynomially many - potential proofs of different
lengths and each length requires its own bit of advice. Algorithm $A$ will use the advice that
collects together all these individual bits.

\subsection{Diagonalization}

Diagonalization in proof complexity was used in 
\cite{Kra-di} (or see \cite[Sec.21.4]{prf}) to prove 
that at least one of the following three statements is true:
\begin{enumerate}

\item There is a function $f : \uu \rightarrow \{0,1\}$
computable in time $2^{O(n)}$ that has circuit complexity $2^{\Omega(n)}$.

\item $\mbox{NP} \neq \mbox{coNP}$.

\item There is no $p$-optimal propositional proof system.

\end{enumerate}
A key part of that is a way, assuming that item 1 fails, how to compress possibly very long proofs and to represent
them by small circuits. Using instead the $Kt$ measure may possibly allow for a stronger result.

\subsection{Random formulas}

M\" uller and Tzameret \cite{MulTza} proved that random 3CNFs with $\Omega(n^{1.4})$ clauses do have
(with the probability going to $1$) polynomial size
refutations in a $TC^0$-Frege system. Their argument is based on formalizing in the proof system 
(via bounded arithmetic) the soundness of the unsatisfiability witnesses proved to exists with a high probability
by Feige, Kim and Ofek \cite{FKO}. 

Such a formula $\tau$ has bit size $m = O(n^{1.4} \log n)$ (and,
by virtue of being random, it has Kt-complexity $\Omega(m)$). 
Feige, Kim and Ofek \cite{FKO} proved that their witness (i.e. also the p-size $TC^0$-Frege proof $\pi$
from \cite{MulTza}) can be found in time $2^{O(n^{0.2}\log n)}$ which is exponential in $m^{\Omega(1)}$. That is,
we know that
\begin{equation}
i_{TC^0-F}(\tau) \le m^{\Omega(1)} \ .
\end{equation}
This leaves open the possibility that this inequality cannot be significantly improved. In that case the formulas
would be witness for Theorem \ref{13.2.21a} for $TC^0$-Frege systems demonstrating even exponential gap.

\section{Concluding remarks} \label{9.4.21c}

Results in Sections \ref{1} and \ref{2} show that the optimality of proof search algorithms
reduces to p-optimality of proof systems in both quasi-orderings based on time or information, respectively. 
This leaves some room for a totally different definition of a quasi-ordering of proof search algorithms
that is coarser than those studied here and in which there could be an optimal algorithm without
implying also the existence of a p-optimal proof system. On the other hand, the ordering by time of Section \ref{1}
is perhaps so rudimentary that it is the finest one among all sensible quasi-orderings; hence the opposite implication 
ought to hold always.
However, it is our view that - from the point of view of proof complexity -
the situation is clarified and the proof search problem as formulated in
\cite[Sec.21.5]{prf}
is simply the p-optimality problem.

This does not quite dispel the doubts about the $\geq_t$ ordering discussed at the end of Section \ref{1}.
The quasi-orderings considered here are theoretical models of a comparison of proof search algorithms
and have shortcomings in modeling actual comparison of practical algorithms that are, we think, quite
analogous to shortcomings of p-time algorithms as a theoretical model of practical feasible algorithms.
The comparison of real life algorithms is also more purpose specific and classifying all purposes
that arise in practice may not be theoretically possible or useful.

However, measure 
$i_P(\tau)$ may still have some uses for comparing two proof systems from the practical proof search point of view. 
For example, it can be used to kill all uniform formulas when testing algorithms (cf. the discussion
at the end of Section \ref{1}) by accepting as test formulas only those satisfying, say, 
$Kt(\tau) \geq (\log |\tau|)^2$. Also, the information-efficiency functions for $P$, $Q$ such that $P >_p Q$
could lead to a suitable distance function measuring how much better $P$ than $Q$ is, by counting
how much more information $Q$-proofs require than $P$-proofs do.

\bigskip

\noindent
{\large {\bf Acknowledgments:}} 

I thank Igor C. Oliveira (Warwick U.) for a discussion about an early draft of the paper.

\end{document}